\documentclass[prl,10pt, twocolumn, nofootinbib, longbibliography]{revtex4-1}

\usepackage{graphicx,amsmath,amssymb}
\usepackage{color}
\usepackage{ulem}
\usepackage{epstopdf}
\usepackage{lmodern}
\usepackage{bm}           
\usepackage{bbm}

\setcitestyle{super,longbibliography}

\def\Rb{$^{87}$Rb }

\def\EE#1{\times 10^{#1}}
\def\ket#1{\left|#1\right\rangle}

\def\ketF#1{\left(#1\right)}

\newcommand{\Eins}{\ensuremath{\mathbbm 1}}

\newcommand{\tr}{{\rm Tr}}

\newcommand{\vect}[1]{\bm{#1}}

\newcommand{\be}{\begin{equation}}
\newcommand{\ee}{\end{equation}}

\newcommand{\beq}{\begin{eqnarray}}
\newcommand{\eeq}{\end{eqnarray}}

\newcommand{\ops}{ {a}_{\rm A}}      
\newcommand{\opi}{ {a}_{\rm B}}       
\newcommand{\opp}{ {a}_{0}}            
\newcommand{\cgamma}{c}
\newcommand{\sgamma}{s}

\begin{document}

\hbadness = 10000

\title{Satisfying the Einstein-Podolsky-Rosen criterion with massive particles}

\author{J. Peise$^1$, I. Kruse$^1$, K.~Lange$^1$, B. L\"u{}cke$^1$, L.~Pezz\`e$^2$, J.~Arlt$^3$, W. Ertmer$^1$, K. Hammerer$^4$, L. Santos$^4$, A.~Smerzi$^2$, C. Klempt$^{1*}$ 
}

\affiliation{$^1$Institut f\"ur Quantenoptik, Leibniz Universit\"at Hannover, Welfengarten~1, D-30167~Hannover, Germany \\
$^2$ QSTAR, INO-CNR and LENS, Largo Enrico Fermi 2, I-50125, Firenze, Italy\\
$^3$ Institut for Fysik og Astronomi, Aarhus Universitet, Ny Munkegade 120, DK-8000 \AA{}rhus C, Denmark\\
$^4$ Institut f\"ur Theoretische Physik, Leibniz Universit\"at Hannover, Appelstra\ss{}e~2, D-30167~Hannover, Germany\\
$^*$ Correspondence should be addressed to C.K. (email:klempt@iqo.uni-hannover.de)
}

\maketitle
\textbf{In 1935, Einstein, Podolsky and Rosen (EPR) questioned the completeness of quantum mechanics by devising a quantum state of two massive particles with maximally correlated space and momentum coordinates. The EPR criterion qualifies such continuous-variable entangled states, where a measurement of one subsystem seemingly allows for a prediction of the second subsystem beyond the Heisenberg uncertainty relation. Up to now, continuous-variable EPR correlations have only been created with photons, while the demonstration of such strongly correlated states with massive particles is still outstanding. Here, we report on the creation of an EPR-correlated two-mode squeezed state in an ultracold atomic ensemble. The state shows an EPR entanglement parameter of $0.18(3)$, which is $2.4$ standard deviations below the threshold $1/4$ of the EPR criterion. We also present a full tomographic reconstruction of the underlying many-particle quantum state. The state presents a resource for tests of quantum nonlocality and a wide variety of applications in the field of continuous-variable quantum information and metrology.}

\begin{figure}[ht!]
	\centering
\includegraphics[width=100mm]{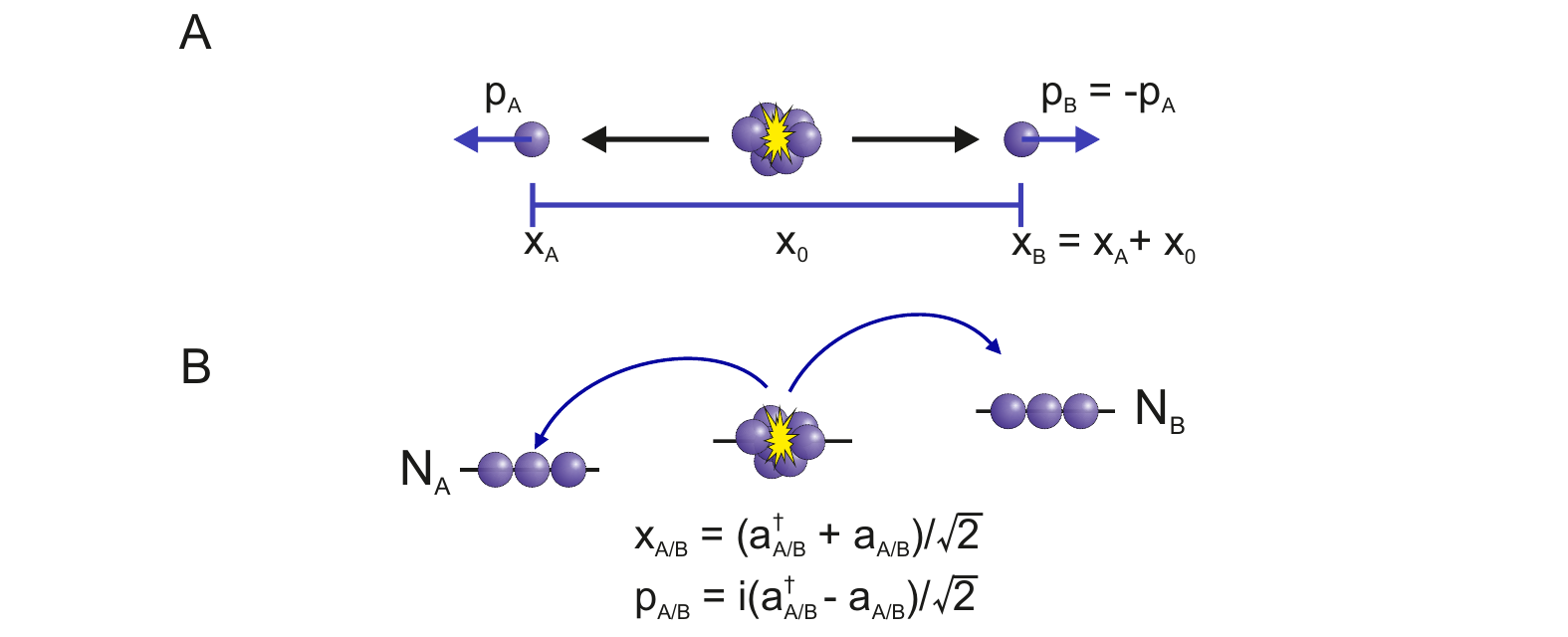}
	\caption{
		\label{fig1}		
			\textbf{Einstein-Podolsky-Rosen correlations. a} EPR's original work describes two particles A and B with maximally correlated position and momentum coordinates $x_\mathrm{A/B}$ and $p_\mathrm{A/B}$. \textbf{b} Spin dynamics in a Bose-Einstein condensate can be used to create EPR correlations between $N_\mathrm{A/B}$ atoms in two different Zeeman levels A and B. The correlations appear in amplitude $x_\mathrm{A/B}$ and phase $p_\mathrm{A/B}$ quadratures which are defined as a function of the creation and annihilation operators $a^\dag_\mathrm{A/B}$ and $a_\mathrm{A/B}$ in the two modes.
	}
\end{figure}

In their original publication~\cite{Einstein1935}, Einstein, Podolsky and Rosen describe two particles A and B with correlated position $x_\mathrm{B}=x_\mathrm{A}+x_0$ and anti-correlated momentum $p_\mathrm{B}=-p_\mathrm{A}$ (see Fig.~\ref{fig1}a). When coordinates $x_\mathrm{A}$ and $p_\mathrm{A}$ are measured in independent realizations of the same state, the correlations allow for an exact prediction of $x_\mathrm{B}$ and $p_\mathrm{B}$. EPR assumed that such exact predictions necessitate an "element of reality" which predetermines the outcome of the measurement. Quantum mechanics however prohibits the exact knowledge of two noncommuting variables like $x_\mathrm{B}$ and $p_\mathrm{B}$, since their measurement uncertainties are subject to the Heisenberg relation $\Delta x_\mathrm{B} \Delta p_\mathrm{B} \geq \frac{1}{2}$. 
EPR thus concluded that quantum mechanics is incomplete - under their assumptions which are today known as "local realism". Later, the notion of EPR correlations was generalized to a more realistic scenario, yielding a criterion~\cite{Furry1936,Reid1989} for the uncertainties $\Delta x_\mathrm{B}^{\mathrm{inf}}$, $\Delta p_\mathrm{B}^{\mathrm{inf}}$ of the inferred predictions for $x_\mathrm{B}$ and $p_\mathrm{B}$. The EPR criterion is met if these uncertainties violate the Heisenberg inequality for the inferred uncertainties $\Delta x_\mathrm{B}^{\mathrm{inf}} \, \Delta p_\mathrm{B}^{\mathrm{inf}} \geq \frac{1}{2}$. The EPR criterion also certifies steering, a concept termed by Schr\"odinger~\cite{Schroedinger1935,Schroedinger1936} in response to EPR which has attracted a lot of interest in the past years~\cite{Wiseman2007}. An experimental realization of states satisfying the EPR criterion is not only desirable in the context of the fundamental questions raised by EPR, but also provides a valuable resource for many quantum information tasks, including dense coding, quantum teleportation~\cite{Braunstein2005} and quantum metrology~\cite{Pezze2014}. Some quantum information tasks specifically require the strong type of entanglement that is tested by the EPR criterion, as for example one-sided device independent entanglement verification~\cite{Opanchuk2014}.

\begin{figure*}[ht!]
	\centering
	\includegraphics[width=0.8\textwidth]{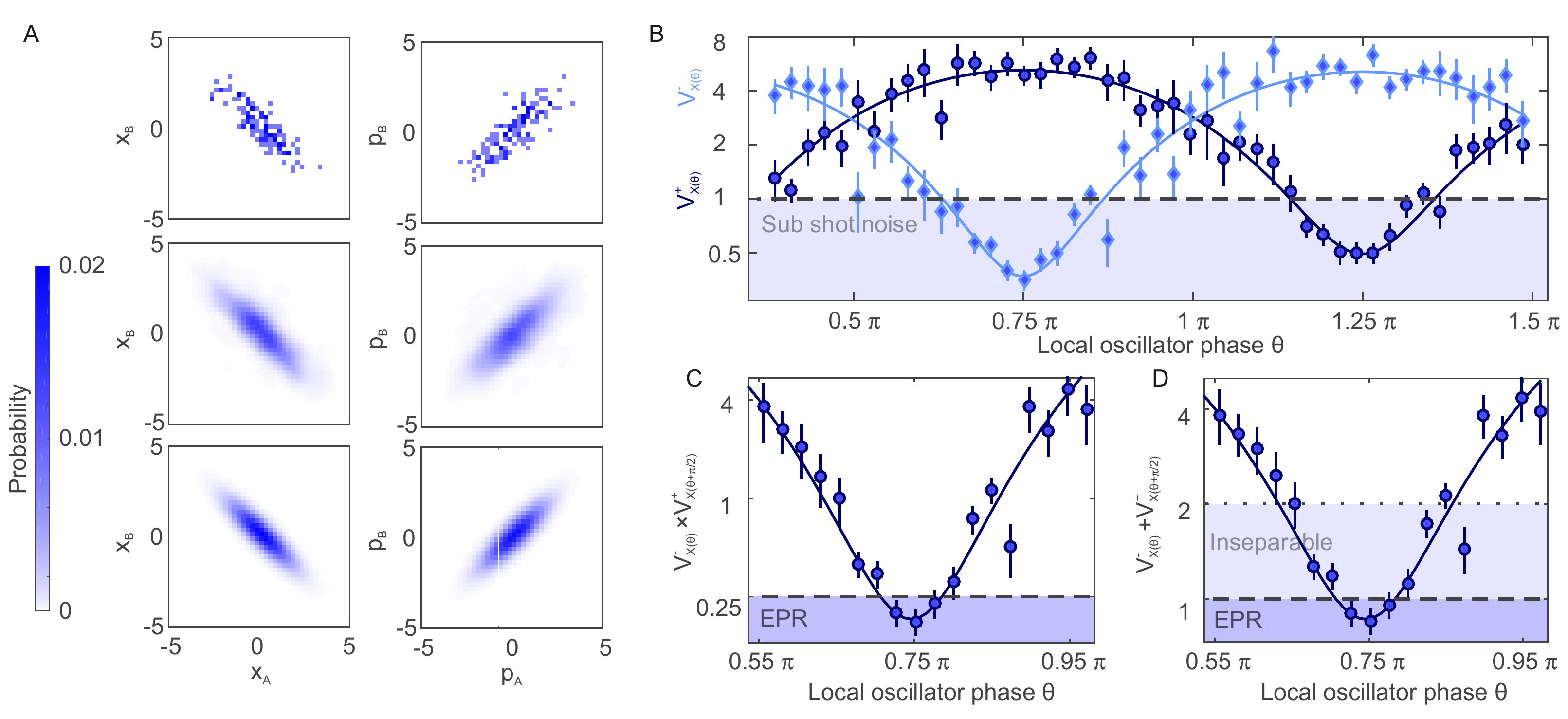}
	\caption{
		\label{fig2}		
		\textbf{Quadrature distributions. a} Recorded probability distributions of the quadratures (first row). Distributions of the quadratures according to the reconstructed state (second row, see below for details on reconstruction). Ideal distributions of the quadratures of a two-mode squeezed state with the reconstructed squeezing parameter $\xi_{\mathrm{fit}} = 0.63$ (third row). \textbf{b} Two-mode variances $V_{\mathrm{X}(\theta)}^+$ and $V_{\mathrm{X}(\theta)}^-$ as a function of the local oscillator phase $\theta$. \textbf{c} EPR parameter $V_{\mathrm{X}(\theta)}^+ \times V_{\mathrm{X}(\theta+\frac{\pi}{2})}^-$ as a function of the local oscillator phase $\theta$. Data points below the dashed line meet the EPR criterion and therefore violate the Heisenberg inequality for inferred uncertainties. At the same time, the EPR criterion also certifies inseparability~\cite{Reid2009}. \textbf{d} The weaker inseparability parameter $V_{\mathrm{X}(\theta)}^+ + V_{\mathrm{X}(\theta+\frac{\pi}{2})}^-$ as a function of the local oscillator phase $\theta$. The dotted line indicates inseparability of the underlying quantum state. The dashed line is a sufficient condition for the EPR criterion. The error bars indicate the statistical uncertainty (one standard deviation).}
\end{figure*}

Up to now, the creation of continuous-variable entangled states satisfying the EPR criterion was only achieved in optical systems. In a seminal publication~\cite{Ou1992}, the EPR criterion was met by a two-mode squeezed vacuum state generated by optical parametric down-conversion. In this experiment, and in more recent investigations~\cite{DAuria2009,Reid2009}, continuous  variables are represented by amplitude $x_\mathrm{A/B}$ and phase $p_\mathrm{A/B}$ quadratures, satisfying the commutation relation $[x_\mathrm{A/B}, p_\mathrm{A/B}]=i$. These quadratures can be measured accurately by optical homodyning. The correlations are captured by the four two-mode variances $V_x^\pm=\mathrm{Var}(x_\mathrm{A} \pm x_\mathrm{B})$ and $V_p^\pm=\mathrm{Var}(p_\mathrm{A} \pm p_\mathrm{B})$. 
These variances were proven to fulfill a symmetric form of Reid's inequality~\cite{Reid1989} $V_x^- \times V_p^+ < \frac{1}{4}$, which is a sufficient EPR criterion since $V_x^- = (\Delta x_\mathrm{B}^{\mathrm{inf}})^2$ and $V_p^+ = (\Delta p_\mathrm{B}^{\mathrm{inf}})^2$. In recent years, continuous-variable entangled optical states have been applied for proof-of-principle quantum computation and communication tasks~\cite{Braunstein2005}. Despite these advances with optical systems, an experimental realization of EPR correlations with massive particles is desirable, because of the similarity to the original EPR proposal and since massive particles may be more tightly bound to the concept of local realism~\cite{Furry1936,Reid1989}.

Entangled states of massive particles have been generated in neutral atomic ensembles, promising fruitful applications in precision metrology due to the large achievable number of entangled atoms~\cite{Esteve2008,Wasilewski2010,Berrada2013,Luecke2014}. They have been created by atom-light interaction at room-temperature~\cite{Julsgaard2001, Wasilewski2010}, in cold samples~\cite{Appel2009,Chen2011,Schleier-smith2010,Haas2014,Behbood2014}, and by collisional interactions in Bose-Einstein condensates~\cite{Esteve2008,Riedel2010,Hamley2012,Luecke2014,Strobel2014}.  For Gaussian states of two collective atomic modes, the inseparability criterion~\cite{Duan2000a,Simon2000} $V_x^- + V_p^+ < 2$ has been used to demonstrate entanglement~\cite{Julsgaard2001, Wasilewski2010,Gross2011}, but the strong correlations necessary to meet the more demanding EPR criterion $V_x^- \times V_p^+ < \frac{1}{4}$ have not been achieved so far.

Here we report on the creation of an entangled state from a spinor Bose-Einstein condensate (BEC) which meets the EPR criterion. We exploit spin-changing collisions to generate a two-mode squeezed vacuum state in close analogy to optical parametric down-conversion. The phase and amplitude quadratures are accessed by atomic homodyning. Their correlations yield an EPR entanglement parameter of $0.18(3)$, which is $2.4$ standard deviations below the threshold $1/4$ of the EPR criterion. Finally, we deduce the density matrix of the underlying many-particle state from a Maximum Likelihood reconstruction.

\section{Results}
\textbf{Two-mode squeezed vacuum.} In our experiments, a BEC with $2 \EE{4}$ \Rb atoms in the Zeeman level $\ketF{F,m_F}=\ketF{1,0}$ generates atom pairs in the levels $\ketF{1,\pm 1}$ due to spin-changing collisions (see Fig.~\ref{fig1}b), ideally yielding the two-mode squeezed state

\begin{equation}
\ket{\xi}=\sum_{n=0}^{\infty} \frac{(-i \tanh{\xi})^n}{\cosh{\xi}} \ket{n,n},
\label{eq:squeezedvac}
\end{equation}

\noindent where $\xi=\Omega t$ is the squeezing parameter, which depends on the spin dynamics rate $\Omega= 2 \pi \times 5.1$~Hz and the spin dynamics duration $t=26$~ms. The notation $\ket{n,m}$ represents a two-mode Fock state in the two Zeeman levels $\ketF{1,\pm 1}$. The generated two-mode squeezed state can be characterized by the quadratures $x_\mathrm{A/B}= \frac{1}{\sqrt{2}} (a^\dag_\mathrm{A/B} + a_\mathrm{A/B})$ and $p_\mathrm{A/B}= \frac{i}{\sqrt{2}} (a^\dag_\mathrm{A/B} - a_\mathrm{A/B})$ for the two levels $\ketF{1,\pm 1}$. These exhibit EPR correlations, since the variances $V_x^- = V_p^+ = e^{-2\xi}$ are squeezed, while the conjugate variances $V_x^+ = V_p^- = e^{2\xi}$ are anti-squeezed. 
The state fulfills Reid's EPR criterion for $\xi>\frac{1}{2} \ln (2)\approx 0.35$ which corresponds to a spin dynamics duration of more than $11$~ms. 
In the limit of large squeezing, our setup presents an exact realization of the perfect correlations with massive particles envisioned by EPR.

\textbf{Quadratures and the EPR criterion.} The quadratures in the two modes are simultaneously detected in our experiments by unbalanced homodyne detection (see methods). Atomic homodyne detection was first demonstrated in Ref.~\citenum{Gross2011}, and later applied to discriminate between vacuum and few-atom states in a quantum Zeno scenario~\cite{Peise2015}.
A small radio-frequency pulse couples $15\%$ of the BEC in the level $\ketF{1,0}$ (the local oscillator) symmetrically to the two modes $\ketF{1,\pm 1}$.
The local oscillator phase is represented by the BEC phase relative to the phase sum of the two ensembles in $\ketF{1,\pm 1}$. It can be varied in our experiments by shifting the energy of the level $\ketF{1,-1}$ for an adjustable time. From the measured number of atoms in both levels, we obtain a linear combination of the quadratures according to $X_\mathrm{A/B}(\theta)= x_\mathrm{A/B} \cos (\theta-\frac{\pi}{4}) +  p_\mathrm{A/B} \sin (\theta-\frac{\pi}{4})$. Figure~\ref{fig2}a shows two-dimensional histograms of these measurements for $\theta = \frac{3}{4} \pi$ and $\theta = \frac{5}{4} \pi$, corresponding to the $x$- and $p$-quadratures. The histograms demonstrate the strong correlation and anticorrelation of these two quadratures, as expected for the EPR case. The variances along the two diagonals, represented by $V_{\mathrm{X}(\theta)}^\pm=\mathrm{Var}(X_\mathrm{A}(\theta) \pm X_\mathrm{B}(\theta))$, are shown in Fig.~\ref{fig2}b and reveal the expected two-mode squeezing behavior. 
From these measurements, we quantify the EPR entanglement by Reid's criterion, yielding $V_x^+ \times V_p^- = 0.18(3)$, which is $2.4$ standard deviations below the limit of $\frac{1}{4}$. 
In addition, the data also fulfills the inseparability criterion as $V_x^+ + V_p^- = 0.85(8)$, which is $15$ standard deviations below the classical limit of $2$ (see Fig.~\ref{fig2}d), and meets the criterion for a symmetric ("two-way") steering between the systems~\cite{Wiseman2007}.
We estimate that the product value could be reduced to $V_x^+ \times V_p^- = 0.13(3)$ if the radio frequency intensity noise was eliminated by stabilization or postcorrection. 
The experimental creation of entangled massive particles which satisfy the continuous-variable EPR criterion presents the main result of this publication.

\begin{figure}[ht!]
	\centering
	\includegraphics[width=\columnwidth]{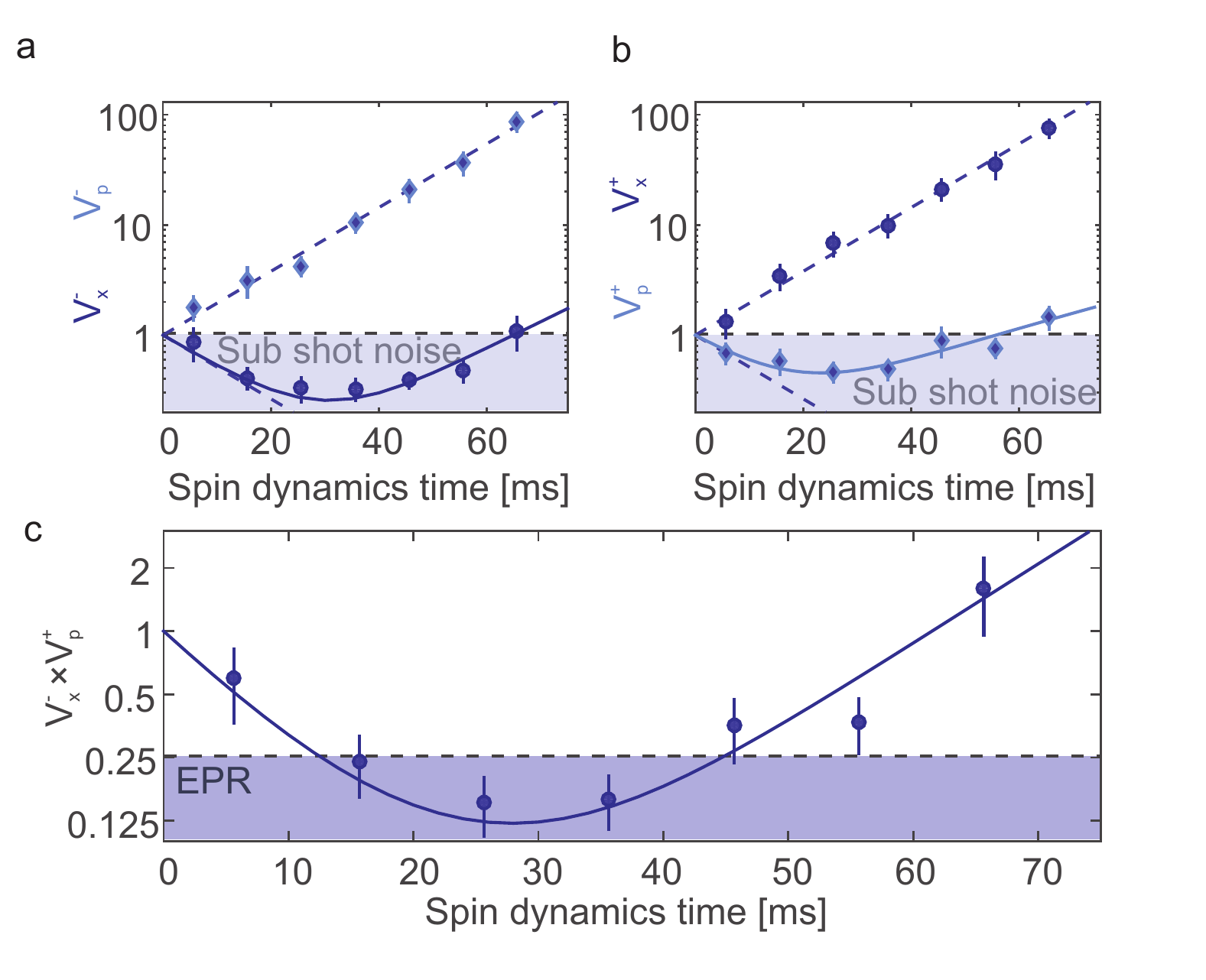}
	\caption{
		\label{fig3}
	\textbf{Squeezing dynamics. a/b} Time evolution of the four two-mode quadrature variances $V_x^\pm$, $V_p^\pm$ during spin dynamics. The data follows the ideal squeezing/anti-squeezing behavior according to the independently measured spin dynamics rate (dashed line). \textbf{c} Time evolution of the EPR parameter $V_x^- \times V_p^+$ during spin dynamics. The squeezed variances and the EPR parameter are well reproduced by a simple noise model (solid line). The error bars indicate the statistical uncertainty (one standard deviation).}
\end{figure}

\begin{figure}[ht!]
	\centering
	\includegraphics[width=\columnwidth]{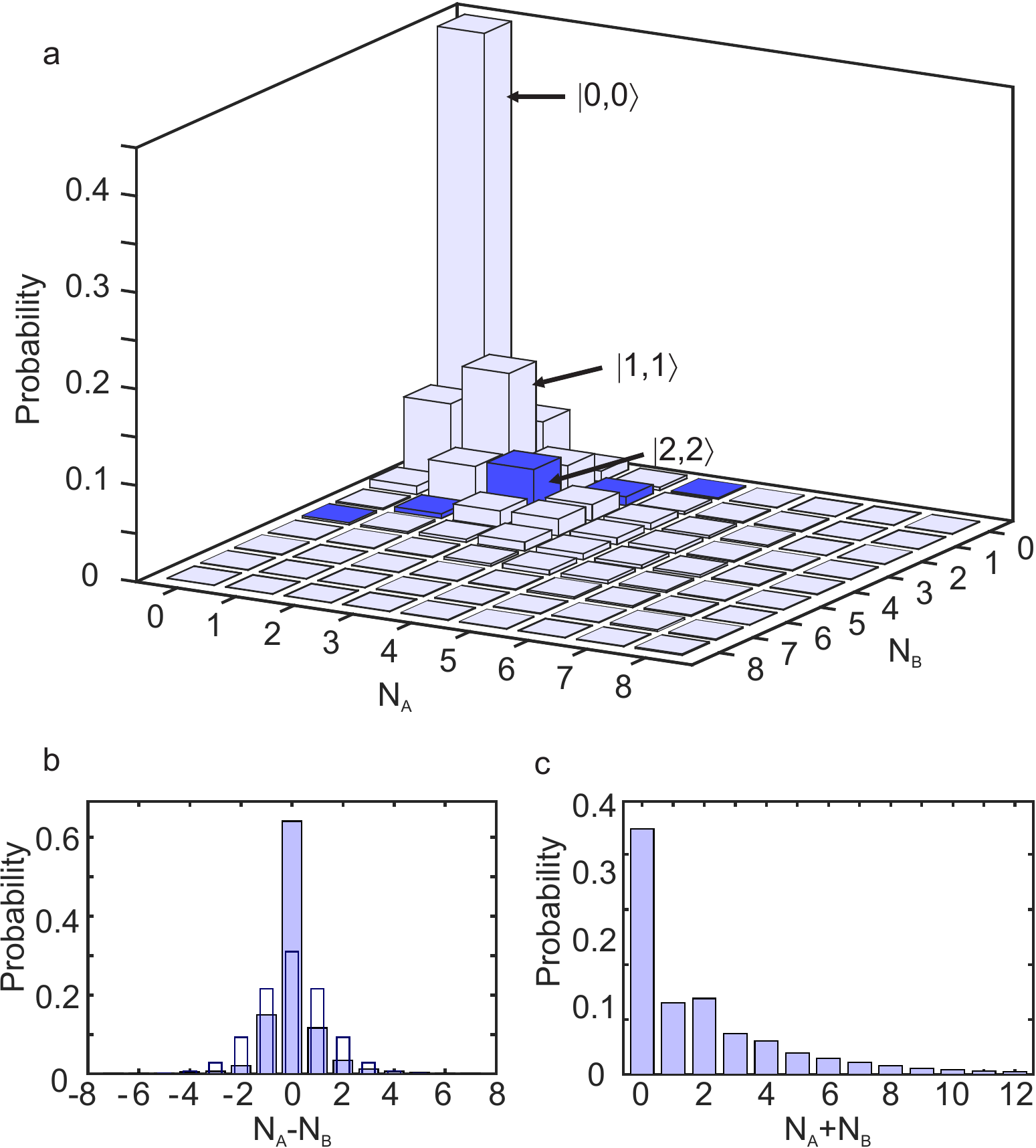}
	\caption{
		\label{fig4}		
		\textbf{Result of the state reconstruction. a} Elements $ \left\langle N_\mathrm{A}, N_\mathrm{B} \right| \rho \left| N_\mathrm{A}, N_\mathrm{B} \right\rangle $ of the reconstructed density matrix. The vacuum state $\ket{0,0}$ and the two-particle twin Fock state $\ket{1,1}$ show the largest values. The large contributions of the twin Fock states reveal the creation of atomic pairs during spin dynamics. For example, the state $\ket{2,2}$ has by far the strongest weight among all states with a total of four particles (highlighted in dark blue).
		\textbf{b} Reconstructed distribution of the difference of the number of atoms in the two modes A and B (solid bars). The distribution is strongly peaked in comparison to a Poissonian distribution with the same mean number of particles (open bars). \textbf{c} Reconstructed distribution of the total number of atoms in the two modes A and B.}
\end{figure}

\textbf{Squeezing dynamics.} Figure~\ref{fig3} shows the squeezing dynamics due to the spin-changing collisions. For these measurements, we fix the local oscillator phase to the values $\theta \approx 3 \pi/4$ and $\theta \approx 5\pi/4$ to record only the $x$- and $p$-variances. As a function of the evolution time, the variances $V_x^-,V_p^+$ are squeezed below the vacuum reference of $1$, while the variances $V_x^+,V_p^-$ exhibit an antisqueezing behavior (Fig.~\ref{fig3}a and b). From these data, we extract the EPR parameter $V_x^- \times V_p^+$,  as a function of evolution time (see Fig.~\ref{fig3}c). The EPR parameter is quickly pushed below $1$ and follows the prediction for an ideal squeezed state. It eventually reaches a minimum at the optimal squeezing time of $26$~ms, as used for the data in Fig.~\ref{fig2}. The data is well reproduced by a simple noise model, which includes a radio-frequency intensity noise of $0.4\%$ and a local oscillator phase noise of $0.044 \pi$ (see methods).

\textbf{Full state reconstruction.} The total of $2864$ homodyne measurements obtained for different local oscillator phases at the optimal evolution time allow 
for a full reconstruction of the underlying many-particle state. 
Previously, tomography of an atomic state was demonstrated either by reconstruction of the Wigner function~\cite{Schmied2011} or the Husimi $Q$-distribution~\cite{Strobel2014,Haas2014}. However, both methods yield a characterization of the state's projection on the fully symmetric subspace only. 
The well-developed methods in quantum optics~\cite{Lvovsky2009} allowed for a full reconstruction of an optical two-mode squeezed state by homodyne tomography~\cite{Vasilyev2000,DAuria2009}. 
Despite the beautiful tomography data, the optical state reconstruction assumed either Gaussian states or averaged over all phase relations, such that the coherence properties could not be resolved.

In contrast, we obtain an unbiased, positive semidefinite density matrix by Maximum Likelihood 
reconstruction~\cite{Lvovsky2009,Hradil2004} of the experimental data, 
free of any a-priori hypothesis. This represents the second major result of this publication. 
The recorded data for each local oscillator phase are binned in two-dimensional histograms (see Fig.~\ref{fig2}a) presenting the marginal distributions for the $x_\mathrm{A/B}$ and $p_\mathrm{A/B}$ variables. 
The reconstructed state is the one which optimally reproduces the measured histograms by a superposition of harmonic oscillator wave functions~\cite{Lvovsky2009}. The coefficients of this superposition are estimates of the density matrix elements of the underlying quantum state (see methods). 

Figure~\ref{fig4} shows the result of the reconstruction. The diagonal matrix elements (Fig.~\ref{fig4}a) witness the predominant creation of atom pairs. The two-particle twin Fock state $\ket{1,1}$ shows the strongest contribution besides the vacuum state. Likewise, the twin Fock states $\ket{2,2}$ to $\ket{4,4}$ have the strongest contribution for a given total number of particles. The strong nonclassicality of the reconstructed state becomes also apparent in the distributions of the difference and the sum of the particles (Fig.~\ref{fig4}b and c). The distribution of the number difference is strongly peaked at zero and is much narrower than a Poissonian distribution with the same number of particles. The distribution of the total number of atoms shows an indication of the characteristic even/odd oscillations, which is caused by the pair production in the underlying spin dynamics. 

\section{Discussion}
For an evaluation of the created state, we have extracted a logarithmic negativity of $1.43 \pm 0.06$ from the reconstructed density matrix. This value is above the threshold of zero for separable states and signals nonseparability of the reconstructed state. 
The quantum Fisher information~\cite{Pezze2009} $F_Q$ for the state projected on fixed-N subspaces reveals that 
$F_Q/\bar{n} = 1.5 \pm 0.1$, where $\bar{n}$ is the average number of particles. 
Since $F_Q/\bar{n}>1$ the state is a resource for quantum enhanced metrology~\cite{Pezze2009}. 
Furthermore,
we fit an ideal two-mode squeezed state with variable squeezing parameter $\xi$ to the reconstructed two-mode density matrix with maximal fidelity. With a fidelity of $78.4$\%, the experimentally created state matches a two-mode squeezed state with a squeezing parameter of $\xi_{\mathrm{fit}}=0.63$. 
The fidelity increases to $90$\% if we include local oscillator phase noise and statistical noise.
The unwanted contributions in the density matrix, including the off-diagonal terms in Fig.~\ref{fig4}a, can be well explained by four effects. 
Firstly, the purity of the reconstructed state is limited by the finite number of homodyne measurements. Secondly, small drifts in the microwave intensity of the dressing field (on the order of $0.1$~\%), which shifts the level $(1,-1)$, result in a small drift of the local oscillator phase. Thirdly, a small drift of the radio-frequency coupling strength during homodyning virtually increases the variance in the $(x_\mathrm{A}+x_\mathrm{B})$ and the $(p_\mathrm{A}+p_\mathrm{B})$ directions. Finally, we did not correct for the detection noise of our absorption imaging.

Our experimental realization of the EPR criterion demonstrates a strong form of entanglement intrinsically connected to the notion of local realism. In the future, the presented atomic two-mode squeezed state allows to demonstrate the continuous-variable EPR paradox with massive particles. Since the two modes A and B are Zeeman levels with an opposite magnetic moment, the modes can be easily separated with an inhomogenous magnetic field to ensure a spatial separation. The nonlocal EPR measurement could then be realized by homodyning with two spatially separated local oscillators. These can be provided by splitting the remaining BEC into the levels $\ketF{2,\pm 1}$ which have the same magnetic moment as the two EPR modes. Furthermore, this setup can be complemented by a precise atom number detection to demonstrate a violation of a Clauser-Horne-Shimony-Holt-type inequality.
Such a measurement presents a test of local realism with continuous-variable entangled states. In this context, neutral atoms provide the exciting possibility to investigate the influence of increasingly large particle numbers and possible effects of gravity.

\section{Methods}
\begin{scriptsize}
\textbf{\footnotesize Experimental sequence.} We start the experiments with an almost pure Bose-Einstein condensate of $20,000$ $^{87}$Rb atoms in an optical dipole potential with trap frequencies of 
$2 \pi \times (280,220,180)$~Hz. At a homogeneous magnetic field of $2.6$~G with fluctuations of about $70$ $\mu$G, the condensate is 
transferred from the level $\ketF{F,m_F}=\ketF{2,2}$ to the level $\ketF{1,0}$ by a series of three resonant microwave pulses. 
During this preparation, two laser pulses resonant to the $F=2$ manifold rid the system of atoms in unwanted spin states. 
Directly before spin dynamics is initiated, the output states $\ketF{1,\pm 1}$ are emptied with a pair of microwave $\pi$-pulses from $\ketF{1,+1}$ to $\ketF{2,+2}$ and from $\ketF{1,-1}$ to $\ketF{2,-2}$ followed by another light pulse. This cleaning procedure ensures that no thermal or other residual excitations are present in the two output modes, which may destroy the EPR signal~\cite{Lewis-Swan2013}.

\begin{figure*}[ht!]
	\centering
	\includegraphics[width=\textwidth]{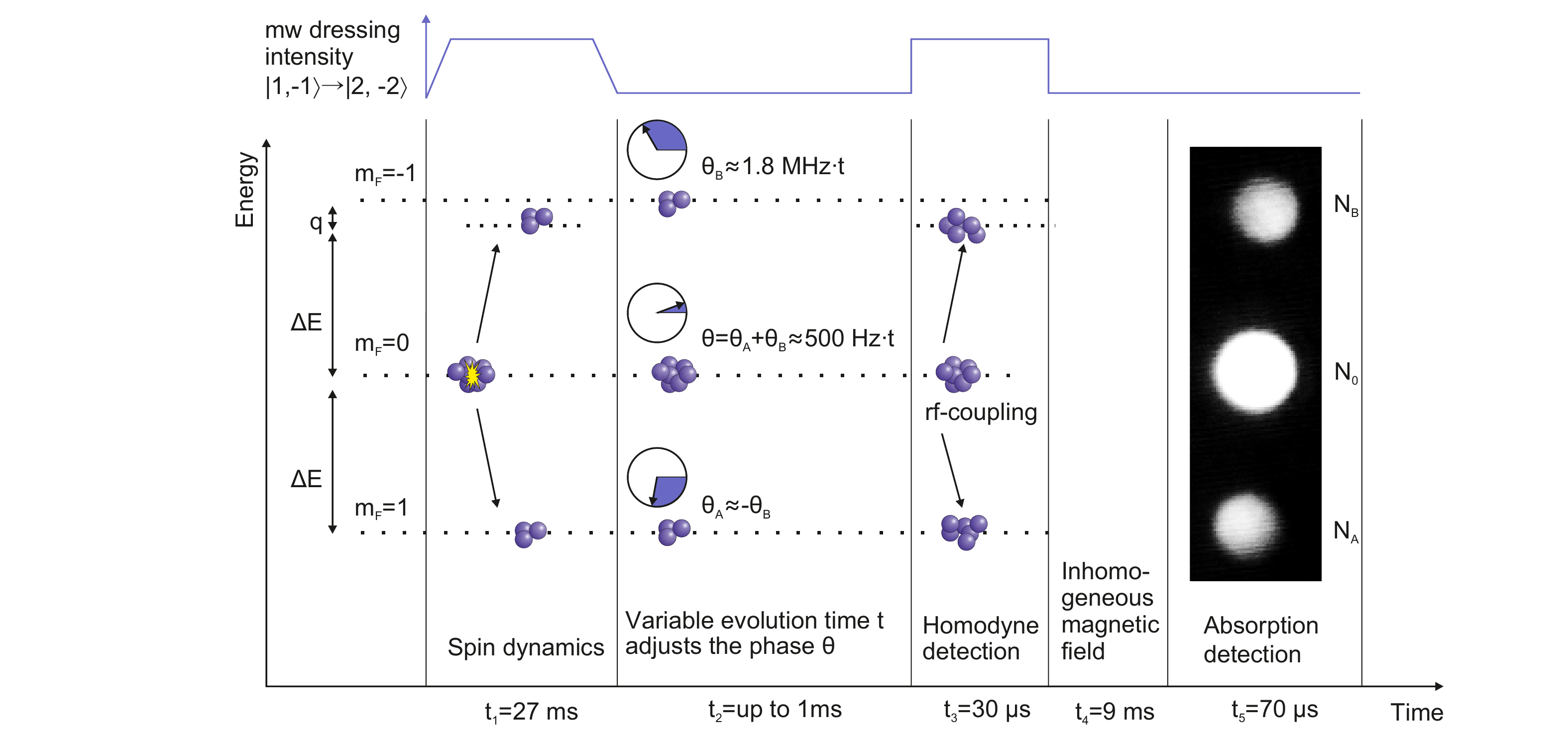}
	\caption{\textbf{Schematic overview of the experimental sequence.} The state is produced via a spin dynamics process at time $t_1$. The remaining condensate in the $m_F=0$ state acts as the local oscillator and has a phase imprinted upon it by an energy shift of the $m_F=-1$ state relative to the $m_F=1$ state. This phase accumulates during time $t_2$ with $500\,\mathrm{Hz}$. At time $t_3$ a radio-frequency pulse couples the $m_F=0$ condensate to the $m_F=\pm 1$ states. This coupling acts as a three port beam splitter. An inhomogeneous magnetic field is applied at $t_4$ before the absorption detection is performed at $t_5$. A microwave dressing shifts the energy of the $m_F=-1$ state during the spin dynamics time $t_1$ and the radio-frequency coupling $t_3$.  } \label{figS1}
\end{figure*}

Figure~\ref{figS1} shows a schematic overview of the following experimental sequence. A microwave frequency which is red-detuned to the transition between the levels $\ketF{1,-1}$ and $\ketF{2,-2}$ by about $208$~kHz is adiabatically ramped on within $675\,\mu$s. The microwave shifts the level $\ketF{1,-1}$ 
by about $500$~Hz, depending on the chosen detuning, to compensate for the quadratic Zeeman effect of $q=491$~Hz, such that multiple spin dynamics resonances can be addressed~\cite{Klempt2010,Luecke2014}. Each resonance condition belongs to a specific spatial mode of the states $\ketF{1,\pm 1}$  to which the atoms are transferred. If the energy of the level $\ketF{1,-1}$ is reduced, more internal energy is released, and higher excited spatial modes are populated (for details, see Ref.~\citenum{Klempt2010}). Here, we choose the first resonance, where spin dynamics leads to a population of the levels $\ketF{1,\pm 1}$ in the ground state of the effective potential. This ensures an optimal spatial overlap between the atoms in the three contributing levels. This resonance condition is reached, when the input state (two atoms in the BEC in the level $\ketF{1,0}$ at the energy of the chemical potential) is exactly degenerate with the output state (two atoms in the levels $\ketF{1,\pm 1}$ including dressing, trap energy and mean-field shift). Due to this degeneracy, the phase relation between the initial condensate and the output state stays fixed during the spin dynamics evolution time. For this configuration, we have checked that spin dynamics is the only relevant process which produces atoms in the state $\ketF{1,\pm 1}$ (see Ref.~\citenum{Peise2015}, Fig.~3). Subsequently, the microwave dressing field is ramped down within $675\,\mu$s, stays off for a variable duration between $25$ and $1150\,\mu$s and is quickly switched on again. The variable hold time allows for an adjustment of the local oscillator phase relative to the output state. 

For the atomic homodyning, a radio-frequency (rf) pulse with a frequency of $1.834$~MHz and a duration of 
$\tau = 30\,\mu$s couples the level $\ketF{1,0}$ with the levels $\ketF{1,\pm1}$. The microwave dressing field is chosen such that both rf transitions are resonant but the resonance condition for spin dynamics is not fulfilled. 
Afterwards, the dipole trap is switched off to allow for a ballistic expansion. After an initial expansion of $1.5$~ms to reduce the density, a strong magnetic field gradient is applied to spatially separate the atoms in the three Zeeman levels. Finally, the number of atoms in the three clouds is detected  by absorption imaging on a CCD camera with a large quantum efficiency. The statistical uncertainty of a number measurement is dominated by the shot noise of the photoelectrons on the camera pixels and amounts to $16$~atoms. We estimate the uncertainty of the total number calibration to be less than $1$\%.\\

\textbf{\footnotesize Three-mode unbalanced homodyning.} 
The rf coupling is described by the three-mode unitary operation $e^{-i {H} \tau/\hbar}$, where 
\be
{H} = \frac{\hbar \Omega_{+1}}{2\sqrt{2}} \big( \ops^\dag \opp + \ops \opp^\dag \big) + \frac{\hbar \Omega_{-1}}{2\sqrt{2}} \big( \opi^\dag \opp + \opi \opp^\dag \big),
\nonumber
\ee
and $\Omega_{\pm 1}$ are Rabi frequencies for the $(1,0) \leftrightarrow (1,\pm1) $ transition (in general $\Omega_{+1} \neq \Omega_{-1}$).
To calculate the mode transformation, we use 
$[{H}, \ops] = - \hbar \Omega_{+1} \opp / 2 \sqrt{2}$, $[{H}, \opi] = - \hbar \Omega_{-1} \opp / 2 \sqrt{2}$ and
$[{H}, \opp] = - \hbar (\Omega_{+1} \ops +  \Omega_{-1} \opi ) /2 \sqrt{2}$.
We have 
\be \label{Umatrix} 
\begin{pmatrix}
\ops \\
\opi \\
\opp
\end{pmatrix}_{\rm out} 
 = 
\begin{pmatrix}
\tfrac{ \tilde{\Omega}_{+1}^2 \cgamma + \tilde{\Omega}_{-1}^2}{2} & 
\tfrac{ \tilde{\Omega}_{+1} \tilde{\Omega}_{-1} ( \cgamma -1 ) }{2} &
\tfrac{ \tilde{\Omega}_{+1} \sgamma }{i \sqrt{2}}  \\
\tfrac{ \tilde{\Omega}_{+1} \tilde{\Omega}_{-1} ( \cgamma -1 ) }{2} &
\tfrac{ \tilde{\Omega}_{-1}^2 \cgamma + \tilde{\Omega}_{+1}^2}{2} &
\tfrac{ \tilde{\Omega}_{-1} \sgamma }{i \sqrt{2}}  \\
\tfrac{ \tilde{\Omega}_{+1} \sgamma}{i \sqrt{2}}  &
\tfrac{ \tilde{\Omega}_{-1} \sgamma}{i \sqrt{2}} &
\cgamma
\end{pmatrix} 
\begin{pmatrix}
\ops \\
\opi \\
\opp
\end{pmatrix}, 
\ee
where $\cgamma = \cos ( \Omega \tau/2)$, $\sgamma = \sin( \Omega \tau/2)$, and 
\be
\tilde{\Omega}_{+1} = \frac{\Omega_{+1}}{\Omega}, 
\qquad 
\tilde{\Omega}_{-1} = \frac{\Omega_{-1}}{\Omega},
\qquad {\rm with} \quad  \Omega = \sqrt{\frac{ \Omega_{+1}^2 + \Omega_{-1}^2 }{2} },
\ee
are rescaled Rabi frequencies. 
Below, we illustrate how the measurement of the number of particles in the $m_F = \pm 1$ mode 
after the rf coupling, $N_{\rm A} = (\ops^\dag \ops)_{\rm out}$ and  $N_{\rm B} = (\opi^\dag \opi)_{\rm out}$, gives access to the number conserving quadratures
\be \label{XAPA} 
x_{\rm A,B}  = \frac{ \hat{a}_{\rm A,B}^\dag \opp + \opp^{\dag} \hat{a}_{\rm A,B} }{\sqrt{2 n_0}},
\qquad  
p_{\rm A,B} = \frac{\opp^{\dag} \hat{a}_{\rm A,B} - \hat{a}_{\rm A,B}^{\dag} \opp}{i \sqrt{2 n_{{0}}}},
\ee
where $n_0 = \langle \opp^{\dag} \opp \rangle$ is the average number of particles in the condensate
before homodyne (similarly, $n_{\rm A,B} = \langle \hat{a}_{\rm A,B}^{\dag} \hat{a}_{\rm A,B} \rangle$). 
In our experiment,  $(n_{\rm A} + n_{\rm B} )/n_0 \sim 10^{-4}$. 
We thus neglect fluctuations of the number of particles in the $m_F=0$ mode, 
replacing $ \opp^\dag \opp$ with its mean value $n_0 \approx n_{\rm A } + n_{0} + n_{\rm B} = N_{\rm tot}$. \\

Number difference. 
The quadrature difference can be experimentally obtained by measuring the 
difference of the number of particles in the $\pm 1$ modes. 
From Eq.~(\ref{Umatrix}) we can directly calculate $N_{\rm A} - N_{\rm B}$.
To the leading order in $n_0$, we have 
\beq \label{Eq.Quad1}
\frac{ N_{\rm A} - N_{\rm B} }{\sqrt{\sgamma^2 N_{\rm tot}}}  &\approx & 
\frac{\sgamma \sqrt{N_{\rm tot}} ( \tilde{\Omega}_{+1}^2 - \tilde{\Omega}_{-1}^2  )}{2}  \nonumber \\
&&  + \frac{ \tilde{\Omega}_{+1} \big[ \cgamma(\tilde{\Omega}_{+1}^2 - \tilde{\Omega}_{-1}^2 ) + 2 \tilde{\Omega}_{-1}^2 \big]}{2 }
 p_{\mathrm{A}} \\
&& + 
\frac{   \tilde{\Omega}_{-1} \big[ \cgamma(\tilde{\Omega}_{+1}^2 - \tilde{\Omega}_{-1}^2 ) - 2 \tilde{\Omega}_{-1}^2 \big]}{2}
p_{\mathrm{B}}.  \nonumber
\eeq
Since $\Omega_{+1}$ and $\Omega_{-1}$ only differ by $1.7\,\%$ in our experiments, and  
$\cgamma(\tilde{\Omega}_{+1}^2 - \tilde{\Omega}_{-1}^2 ) \ll 2 \tilde{\Omega}_{\pm1}^2$, 
we can simplify this equation and obtain 
\be \label{QuadratureDif} 
p_{\rm A} - p_{\rm B} =
\frac{ N_{\rm A} - N_{\rm B} - 
	\sgamma^2 \big( \tilde{\Omega}_{+1}^2 - \tilde{\Omega}_{-1}^2 \big) N_\textrm{tot}/2}
	{\sqrt{ \sgamma^2 N_\textrm{tot} }}.
\ee

Number sum. 
The quadrature sum is obtained by adding the number of particles in the $\pm 1$ modes after homodyning: 
\beq
\frac{ N_{\rm A} + N_{\rm B} }{\sqrt{\sgamma^2 \cgamma^2 N_{\rm tot}}}
&\approx& 
\frac{ \sgamma  \sqrt{N_{\rm tot}} }{\cgamma}
+ \tilde{\Omega}_{+1} p_{\rm A} + \tilde{\Omega}_{-1}  p_{\rm B}
\eeq
Taking $\tilde{\Omega}_{-1} \approx \tilde{\Omega}_{+1} \approx 1$, 
we have 
\be \label{QuadratureSum} 
p_{\rm A} + p_{\rm B} =
\frac{ N_{\rm A} + N_{\rm B} - \sgamma^2 N_\textrm{tot}}
	{\sqrt{ \sgamma^2 \cgamma^2 N_\textrm{tot} }}.
\ee

Finally, the mean transfer of particles from $m_F = 0$ to $m_F = \pm 1$ and the mean number difference is used to calculate
\be
\left\langle \frac { N_{\rm A} + N_{\rm B} }{N_\mathrm{tot}} \right\rangle \approx
\sgamma^2, \quad
{\rm and}  
\quad
\left\langle \frac{ N_{\rm A} - N_{\rm B} }{N_\mathrm{tot}} \right\rangle \approx
\frac 1 2 \sgamma^2 \big( \tilde{\Omega}_{+1}^2 - \tilde{\Omega}_{-1}^2 \big). 
\ee
Observing a transfer of {15\%} of the atoms from $m_F=0$ to $m_F=\pm 1$ we deduce $\cgamma^2 \approx 0.85$. \\

To summarize, Eqs.~(\ref{QuadratureDif}) and (\ref{QuadratureSum}) are used to experimentally obtain
$p_{\rm A} \pm p_{\rm B}$ from the measurement of the number of particles in the output modes. 
The quadratures $x_{\rm A} \pm x_{\rm B}$ are obtained with the same method, by applying a relative $\pi/2$
phase between the pump and side modes before homodyne detection.\\

\textbf{\footnotesize Entanglement criteria for continuous variables.}
Criteria for identifying continuous-variable entanglement between the systems A and B, with no 
assumption on the quantum state of the local oscillator, have been discussed in Ref.~\citenum{Ferris2008}. \\

{\it Separability.} For mode-separable states, ${\rho}_{\rm sep} = \sum_k p_k {\rho}^{(k)}_{\rm A} \otimes {\rho}^{(k)}_{\rm B}$
($p_k >0$, $\sum_k p_k =1$), 
we have \cite{Ferris2008, Raymer2003}
\be \label{Eq.separability} 
V_x^{\pm} + V_p^{\mp}
\geq 2 - \frac{ {n}_{\rm A} + n_{\rm B}}{{n}_{0}},
\ee
where $V_x^\pm = {\rm Var} (x_{\mathrm{A}} \pm x_{\mathrm{B}})$ and $V_p^\pm = {\rm Var} (p_{\mathrm{A}} \pm p_{\mathrm{B}})$ are the variances of quadrature sum and difference, respectively.
A violation of Eq.~(\ref{Eq.separability}) signals non-separability, i.e. ${\rho} \neq {\rho}_{\rm sep}$.
Equation (\ref{Eq.separability}) generalizes the criterion of Refs.~\citenum{Duan2000a} and \citenum{Simon2000} that 
was derived for standard quadrature operators
(i.e. when the $m_F=0$ mode is treated parametrically, the operator $\opp$ being replaced by $\sqrt{n_0}$ ).  \\

{\it EPR criterion.} 
Reid's EPR criterion corresponds to a violation of the Heisenberg uncertainty relation on system B, 
when measurements are performed on system A. 
This requires the two-mode state to be non-separable and to have strong correlations between the 
sum and difference of position and momentum quadratures,  
$x_{\rm A} \pm x_{\rm B}$ and $p_{\rm A} \mp p_{\rm B}$.
We point out that not all non-separable states fulfill Reid's criterion. 
The position-momentum quadratures for the B mode satisfy the 
commutation relation 
$[x_{\rm B}, p_{\rm B}] = -i ( \opi^{\dag} \opi - \opp^{\dag} \opp )/ {n}_0$.
The corresponding Heisenberg uncertainty relation is
$ \Delta^2 x_{\rm B}  \Delta^2 p_{\rm B}  \geq \tfrac{1}{4} ( 1 - \tfrac{{n}_{\rm B}}{{n}_0} )^2$.
Let us introduce the quantities $x_{\rm ext}(x_{\rm A})$ and $p_{\rm ext}(p_{\rm A})$, which are 
the estimate of $x_{\rm B}$ and $p_{\rm B}$ on system B, respectively,
given the results of quadrature measurements on the system A.
We then indicate as $\Delta^2 x_{\rm B}^{\rm inf}$ the squared deviation of the estimate from the actual value, 
averaged over all possible results $x_{\rm A}$, 
\be
\Delta^2 x_{\rm B}^{\rm inf} = \int d x_{\rm B} \int d x_{\rm A} P(x_{\rm A}, x_{\rm B}) [x_{\rm B} - x_{\rm ext}(x_{\rm A})]^2, 
\ee 
and similarly for $\Delta^2 p_{\rm B}^{\rm inf}$,
where $P(x_{\rm A}, x_{\rm B})$ is the joint probability.
Reid's criterion thus reads~\cite{Ferris2008}
$\Delta^2 x_{\rm B}^{\rm inf}  \Delta^2 p_{\rm B}^{\rm inf}  < \tfrac{1}{4} ( 1 - \tfrac{{n}_{\rm B}}{{n}_0} )^2$.
Taking $x_{\rm ext}(x_{\rm A}) = x_{\rm A} - (\bar{x}_{\rm A} - \bar{x}_{\rm B})$ and 
$p_{\rm ext}(p_{\rm A}) = - p_{\rm A} + (\bar{p}_{\rm B} + \bar{p}_{\rm A})$, where the bar indicates statistical average, 
Reid's criterion translates into a condition for the product of quadrature variances:
\be \label{Eq.Reid} 
V_x^- \times V_p^+ < \frac{1}{4} \bigg( 1 - \frac{{n}_{\rm B}}{{n}_0} \bigg)^2.
\ee
In our case, ${n}_{\rm A}/{n}_0 \sim {n}_{\rm B}/{n}_0 \approx 10^{-4}$. 
Therefore corrections in Eqs.~(\ref{Eq.separability}) and (\ref{Eq.Reid})
due to finite number of particles in the $m_F=0$ are negligible. 
We are thus in a continuous-variable limit. 

We point out that the above EPR criterion -- consistent with the analysis of the experimental data presented in Figs.~\ref{fig2}b and \ref{fig3} --
uses quadrature variances with symmetric contributions from A and B.
In this case the EPR threshold is 1/4. 
The above inequalities and entanglement criteria can be generalized (and optimized) for asymmetric contributions, see Refs.~\citenum{Reid1989} and \citenum{Reid2009}.\\

\textbf{\footnotesize Quantum state tomography.}
Here we discuss the protocol used for quantum state tomography and, very briefly, its theoretical basis.
A more detailed discussion can be found in Refs. \citenum{Lvovsky2009} and \citenum{Hradil2004}. 
We point out that our state reconstruction is performed without assumption neither on the state nor on the experimental quadrature distribution,
in particular we do not assume our state to be a Gaussian state. 

We have collected a total of $N=2864$ measurements of the quadratures $x_{\rm A}$ and $x_{\rm B}$ at different values of 
the local oscillator phase $\theta$ relative to the side modes. 
The measurement results are binned in 2D histograms 
(see Fig.~2a, where the typical bin width is $dx=0.25$) such that we take $x_{\rm A,B}$ to have a discrete spectrum.
To simplify the notation, let us indicate as $x$ the square bin $[x_{\rm A},x_{\rm A}+dx]$,  $[x_{\rm B}, x_{\rm B}+dx]$.
Given a quantum state ${\rho}$ (unknown here),
the probability to observe a certain sequence of results ($n_{x,\theta}$ measurement in the bin $x$, when the phase value is set to $\theta$, 
with $\sum_{x,\theta} n_{x,\theta} = N$) is
\be \label{likelihood} 
\mathcal{L} (\{ n_{x, \theta} \} \vert {\rho} ) = \frac{N!}{\prod_{x,\theta} n_{x,\theta}!} 
\prod_{x,\theta} P_{{\rho}}(x,\theta)^{n_{x,\theta}},
\ee
indicated as likelihood function. 
In Eq.~(\ref{likelihood}), 
$P_{{\rho}}(x,\theta) = P_{{\rho}} (x \vert \theta)P(\theta) $
is the joint probability, 
$P_{{\rho}} (x \vert \theta)= \langle x \vert U_\theta^\dag {\rho}  U_\theta  \vert x \rangle $
is the conditional probability, with 
$\vert x \rangle = \vert x_{\rm A}, x_{\rm B} \rangle$,
$ U_{\theta} = e^{-i \theta ({n}_A + {n}_B)}$, and $P(\theta)$ is 
the fraction of measurements done when the phase is equal to $\theta$.
The maximum likelihood (ML)
\be
{\rho}_{\rm ML} = \arg \big[ \max_{{\rho}} \mathcal{L} (\{ n_{x, \theta} \} \vert {\rho} ) \big] 
\ee
is the state that maximizes $\mathcal{L} (\{ n_{x, \theta} \} \vert {\rho} )$ on the manifold of density matrices.
To find the ML we use the chain of inequalities~\cite{Lvovsky2009,Hradil2004}
\beq
\mathcal{L} (\{ n_{x, \theta} \} \vert {\rho} )^{\frac{1}{N}} &\leq& 
\tr\big[ {\rho} {R}(\vect{f}, \vect{a}) \big] \prod_{x, \theta} (a_{x, \theta})^{f_{x, \theta}} 
\leq 
\lambda(\vect{f}, \vect{a}) \prod_{x, \theta} (a_{x, \theta})^{f_{x, \theta}}, 
\eeq
where $a_{x, \theta}$ are arbitrary positive numbers ($\vect{a} = \{ a_{x, \theta}\}$ is the corresponding vector),
$f_{x, \theta} = n_{x,\theta}/N$ are relative frequencies ($\vect{f} = \{ f_{x, \theta}\}$ is the corresponding vector), and
\be
{R} =
\frac{1}{N} \sum_{x, \theta} \frac{n_{x, \theta}}{P(x \vert \theta)} U_\theta \vert x \rangle \langle  x \vert U_\theta^\dag, 
\ee
is a non-negative operator with largest eigenvalue $\lambda(\vect{f}, \vect{a}) $. 
The second inequality is saturated by taking 
${\rho}={\rho}_{\rm ML}$ with support 
on the subspace corresponding to the maximum eigenvalue of ${R}$:
${R}(\vect{f}, \vect{a}) {\rho}_{\rm ML} = \lambda(\vect{f}, \vect{a}) {\rho}_{\rm ML}$. 
The first inequality is a Jensen's inequality between the geometric and the arithmetic average 
(which follows from the concavity of the logarithm). It is saturated if and only if  
$a_{x, \theta} = P(x,\theta)$, which also implies 
$\tr[{R}(\vect{f}, \vect{a}) {\rho}_{\rm ML}]=1$ and thus $ \lambda(\vect{f}, \vect{a}) = 1$. 
In conclusion, the search for the ML can be recast in the operator equation  
${R} {\rho}_{\rm ML} = {\rho}_{\rm ML}$ or, equivalently 
(since $ R$ and $ \rho_{\rm ML}$ are Hermitian operators), 
\be \label{EqR} 
{R} {\rho}_{\rm ML} {R} = {\rho}_{\rm ML}. 
\ee
Numerically, this equation is solved iteratively:
we start the protocol from a unit matrix ${\rho}^{(0)} = \mathcal{N}[\Eins]$
and apply repetitive iterations according to Eq.~(\ref{EqR}), 
${\rho}^{(k+1)} = \mathcal{N}[{R}({\rho}^{(k)}) {\rho}^{(k)} {R}({\rho}^{(k)}) ]$
being the $k$th step of the algorithm, where $\mathcal{N}$ denotes normalization to unit trace.
The convergence (which is not guaranteed in general) is checked. 
The method guarantees that ${\rho}_{\rm ML}$ is a non-negative definite operator. 
In practical implementations, it is most convenient to work in the atom-number representation and write 
${\rho} = \sum_{n_{\rm A}, n_{\rm B}, m_{\rm A}, m_{\rm B}=0}^{n_{\rm cut}} \rho_{n_{\rm A}, n_{\rm B}, m_{\rm A}, m_{\rm B}} 
\vert n_{\rm A}, n_{\rm B} \rangle \langle m_{\rm A}, m_{\rm B} \vert$, 
where $n_{\rm cut}$ is a cutoff (in our case $n_{\rm cut} \approx 10$).
We use
$\langle n \vert x_{\rm A,B} \rangle = \tfrac{e^{-x_{\rm A,B}/2}}{\pi^{1/4} \sqrt{ 2^n n ! }} H_n(x_{\rm A,B})$, where
$H_n$ is the Hermite polynomial of order $n$. \\

\textbf{\footnotesize Simulation of ideal state reconstruction.}
To check the consistency of the used tomography method, we have simulated the reconstruction of
an ideal two-mode squeezed vacuum state $\vert \xi \rangle$, Eq.~(\ref{eq:squeezedvac}). 
The simulation follows three steps: 
{\it i)} we generate distributions for the quadratures $x_{\rm A,B}$ at different values of $\theta$,
according to the probability 
$P(x_{\rm A,B} \vert \theta) = \vert \langle \xi \vert {U}_\theta \vert x_{\rm A,B} \rangle \vert^2$;
{\it ii)} we 	generate $p$ random quadrature data for each $\theta$ value (for a total of $N = p \times n_\theta$, where $n_\theta$
is the number of $\theta$ values considered). This simulates, via Monte Carlo sampling, the acquisition of experimental data. 
{\it iii)} We perform a ML reconstruction, using the same numerical code and method
used for the analysis of the experimental data.
In Fig.~\ref{FigS2} we plot the quantum fidelity between the reconstructed state, ${\rho}_{\rm ML}$, 
and the two-mode squeezed vacuum state,
$\mathcal{F} = \sqrt{ \langle \xi \vert {\rho}_{\rm ML} \vert \xi \rangle }$.  
When the number of measurements $p$ per $\theta$ value is increased, the fidelity 
converges to an asymptotic value $\mathcal{F}_{\infty} \lesssim 1$. 
The asymptotic fidelity $\mathcal{F}_{\infty}$ tends to $1$ when decreasing the bin size $dx$.

\begin{figure}[ht!]
\begin{center}
\includegraphics[clip,width=0.8 \columnwidth]{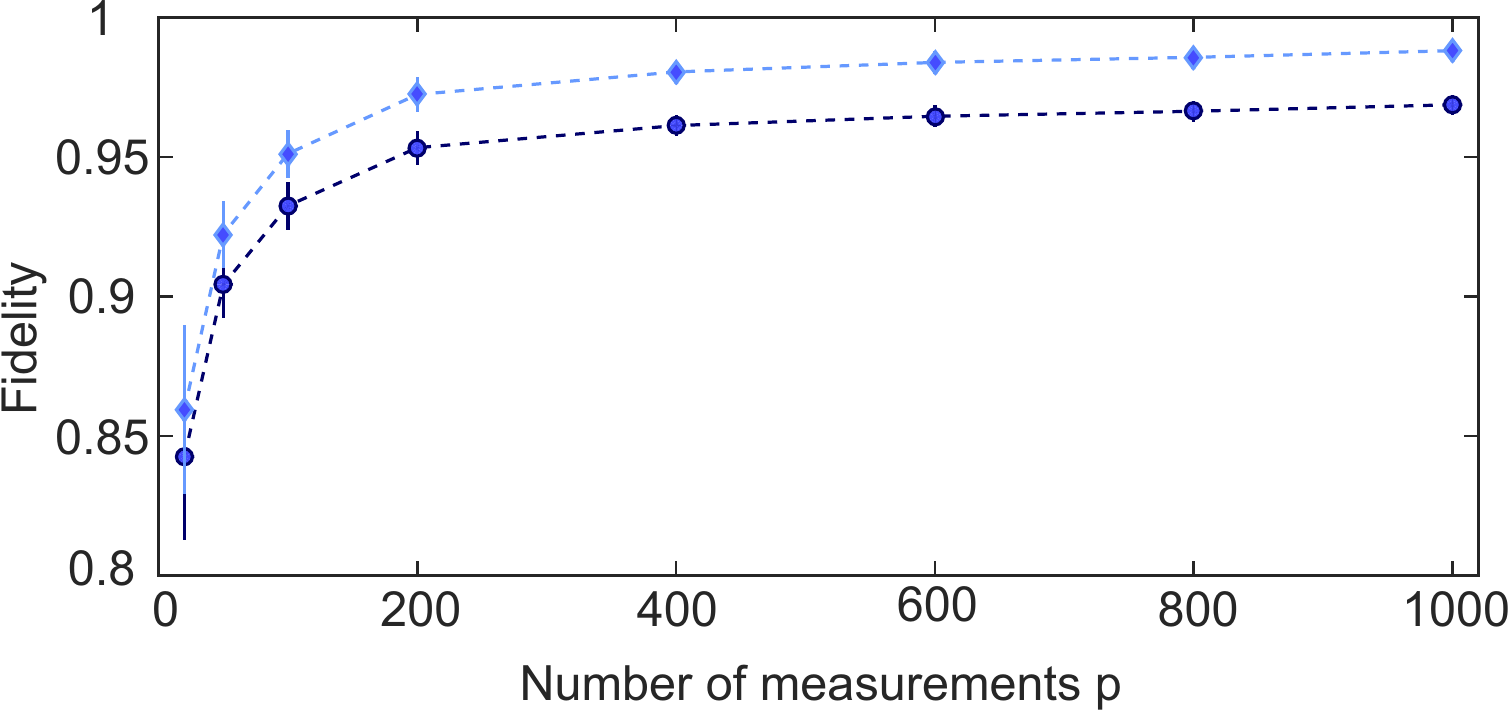}
\end{center}
\caption{\textbf{Simulated quantum state tomography of an ideal squeezed state.}
Here we show the quantum fidelity between the reconstructed state ${\rho}_{\rm ML} $ 
and the ideal two-mode squeezed vacuum state, as a function of the number of measurements $p$ per $\theta$ value
(here we consider $n_\theta=29$ phase values).
The dark blue circles are obtained for a binning of $dx=0.25$, the light blue diamonds for $dx=0.1$. 
The dashed lines are a guide to the eye, the error bars (one standard deviation) are obtained with a bootstrap method.}  \label{FigS2}
\end{figure}

Furthermore, to characterize the entanglement of the reconstructed state, we have evaluated the logarithmic negativity 
and the quantum Fisher information (QFI).
The logarithmic negativity is defined as 
$E[{\rho}] = \log_2 \sqrt{{\rho}_{\rm ppt} {\rho}_{\rm ppt}^\top }$,
where ${\rho}_{\rm ppt}$ is the partial transpose of ${\rho}_{\rm ML}$.
Mode-entanglement is obtained for\cite{Guhne2009} $E[{\rho}] > 0$.
The QFI for the state projected over subspaces of a fixed number of particles $n$, 
$\rho = \sum_n Q_n \rho_n$, is given by~\cite{Pezze2015} 
\be
F_Q[{\rho}] = 2 \sum_n Q_n 
\sum_{k_n, k_n'} \frac{(p_{k_n} - p_{k'_n})^2}{p_{k_n} + p_{k'_n}} \vert \langle k_n \vert {J}_{\vect r} \vert k_n' \rangle \vert^2
\ee
where ${\rho}_n 
= \sum_{k_n} p_{k_n} \vert k_n \rangle \langle k_n \vert$
is in diagonal form 
and ${J}_{\vect r}$ is the collective pseudo-spin operator (pointing along an arbitrary direction $\vect{r}$ in the three-dimensional space).
The QFI is then maximized over $\vect{r}$, for further details see Refs.~\citenum{Pezze2014}.
Particle entanglement, useful for sub-shot-noise metrology, is obtained for\cite{Pezze2015} $F_Q[{\rho}] \geq \bar{n}$, 
where $\bar{n}$ corresponds to the average number of particles in the two-mode state.
Similarly to the results of simulations shown in Fig.~\ref{FigS2} we obtain that, 
in the limit $p \to \infty$ and $dx \to 0$, the logarithmic negativity and QFI converge to 
$E[\vert \xi \rangle] = 2 \xi/\log 2$ and $F_Q[\vert \xi \rangle] = 4 \sinh^2 \xi \cosh^2 \xi$, respectively, 
which are analytical values calculated for the two-mode squeezed vacuum state. \\

\begin{figure}[ht!]
\begin{center}
\includegraphics[clip,width=0.8 \columnwidth]{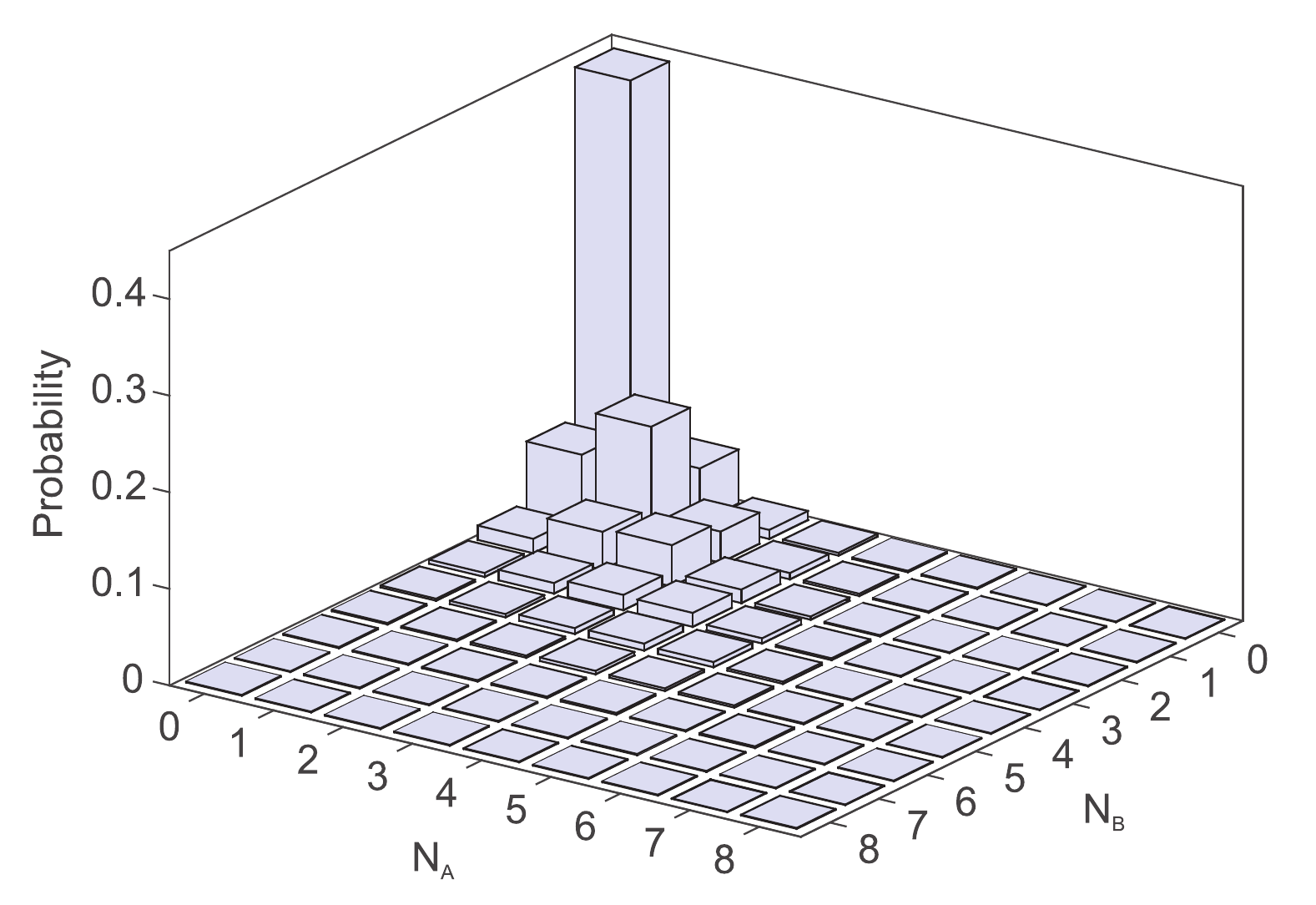}
\end{center}
\caption{\textbf{Simulated quantum state tomography of a noisy squeezed state.}
Tomography reconstruction of a two-mode squeezed state affected by 
phase noise and a systematic shift in the variance of the quadrature sum, as in the experiment.
Here we have used the experimental parameters: $p \approx 100$ measurements, $n_{\theta}=29$ phase values and a bin size of $dx=0.25$. 
The reconstructed state agrees very well with Fig.~4 of our manuscript, both qualitatively 
(showing both the presence of off-diagonal terms and asymmetry) and quantitatively
(with a quantum a fidelity of about 90\%).}  \label{FigS3} 
\end{figure}

\textbf{\footnotesize Noise model and simulation of noisy state reconstruction.}
The main sources of noise in our apparatus are phase fluctuations and noise of the rf coupling strength.
The phase noise is assumed to have a Gaussian distribution 
$P_\sigma(\theta) = e^{- \theta^2/2 \sigma^2}/\sqrt{2 \pi \sigma^2}$ and we estimate a width $\sigma \approx 0.36$.
Correlated fluctuations of $\Omega_{+1}$ and $\Omega_{-1}$ affect (to first order) only measurements of the quadrature sum. 
We have evaluated that this effect systematically increases the variance by $0.12$.
Both these effects are included in the solid line of Fig.~3.

We have simulated the state reconstruction in presence of these noise effects. 
We model the state in presence of phase noise as
\be \label{TMSSPN} 
{\rho}_{\rm pn}(\sigma) = \int_{-\pi}^{\pi} d \theta P_\sigma(\theta) \vert \xi, \theta \rangle  \langle \xi, \theta \vert
\ee
where $\vert \xi, \theta \rangle = \sum_{n=0}^{+\infty} e^{-i n \theta} \frac{(\tanh \xi)^n}{\cosh \xi} \vert n, n \rangle$.
The systematic shift of the quadrature sum is included in the calculation of the quadrature distributions used to generate random data.
Results are shown in Fig.~\ref{FigS3}. 
We see that statistical noise (i.e. the limited sample size) 
and phase noise are responsible for the appearance of off-diagonal terms, 
very similar to the ones observed in Fig.~\ref{fig4}. Note that phase noise alone is not responsible for the appearance of off-diagonal terms in the density matrix. 
This can be seen by rewriting Eq.~(\ref{TMSSPN}) as $ {\rho}_{\rm pn}(\sigma)= \sum_{n,m=0}^{+\infty} \tilde{P}_\sigma(n-m) \tfrac{(\tanh \xi)^{n+m}}{\cosh^2 \xi} \vert n,n \rangle \langle m,m \vert$, where $\tilde{P}_\sigma(n-m) = \int_{-\pi}^{\pi} d \theta P_\sigma(\theta) e^{i (n-m) \theta }$.

Figure \ref{FigS3} shows a slight asymmetry of the reconstructed state due to the systematic shift of the variance sum: 
this effect is also observed in Fig.~\ref{fig4}. The quantitative agreement between the simulated density matrix  $\rho_{\rm sim}$ and the experimental density matrix $\rho_{\rm exp}$ is excellent,
with a quantum fidelity ${\rm Tr}[\sqrt{\sqrt{\rho_{\rm sim}} \rho_{\rm exp} \sqrt{\rho_{\rm sim}}}] \approx 0.9$. 
\end{scriptsize}

\section{Acknowledgments}
\begin{scriptsize} We thank G. T\'o{}th for inspiring discussions. We thank W. Schleich for a review of our manuscript. We acknowledge support from the Centre for Quantum Engineering and Space-Time Research (QUEST) and from the Deutsche Forschungsgemeinschaft (Research Training Group 1729). We acknowledge support from the European Metrology Research Programme (EMRP). The EMRP is jointly funded by the EMRP participating countries within EURAMET and the European Union. L.P. and A.S. acknowledge financial support by the EU-STREP project QIBEC. J.A. acknowledges support by the Lundbeck Foundation.

\end{scriptsize}

\end{document}